# Applied Evaluative Informetrics

Henk F. Moed

**Preprint (author copy) of PART 1 of a book to be published by Springer in the summer of 2017**

# Table of Contents

Part 5. Lectures

13.  From Derek Price's Network of Scientific papers to advanced science mapping

13.1.  Abstract

13.2.  Networks of scientific papers

13.3.  Modelling the relational structure of subject space.

13.4.  Mapping softwares

14.  From Eugene Garfield's Citation index to Scopus and Google Scholar

14.1.  Abstract

14.2.  Science Citation Index and Web of Science

14.3.  Scopus versus Web of Science

14.4.  Google Scholar versus Scopus

14.5.  Concluding remarks

15.  From Francis Narin's science-technology linkages to double boom cycles in technology

15.1.  Abstract

15.2.  Citation analysis of the science-technology interface

15.3.  Theoretical models of the relationship between science and technology

15.4.  Double boom cycles in product development

15.5.  Selected case studies

16.  From journal impact factor to SJR, Eigenfactor, SNIP, CiteScore and Usage Factor

16.1.  Abstract

16.2.  Journal impact factors

16.3.  Effect of editorial self-citations

16.4.  Alternative journal metrics

17.  From relative citation rates to altmetrics

17.1.  Abstract

17.2.  Citation-based indicators

17.3.  Usage-based indicator and altmetrics

17.4.  Efficiency indicators

PART 6. Papers

18.  A comparative study of five world university rankings

# Preface

In 1976, Francis Narin, founder and for many years president of the information company CHI Research, published a seminal report to the US National Science Foundation entitled *Evaluative Bibliometrics: The use of publication and citation analysis in the evaluation of scientific activity*. The current book represents a continuation of his work. It is also an update of an earlier book published by the current author in 2005, *Citation Analysis in Research Evaluation*. In the past 15 years, many new developments have taken place in the field of quantitative research assessment, and the current book aims to describe these, and to reflect upon the way forward.

Research assessment has become more and more important in research policy, management and funding, and also more complex. The role of quantitative information has grown substantially, and research performance is more and more conceived as a *multi-dimensional* concept. Currently not only the classical indicators based on publication and citation counts are used, but also new generations of indicators are being explored, denoted with terms such as altmetrics, webometrics, and usage-based metrics, and derived from multiple multi-disciplinary citation indexes, electronic full text databases, information systems' user log files, social media platforms and other sources. These sources are manifestations of the computerization of the research process and the digitization of scientific-scholarly communication. This is why the current book uses the term *informetrics* rather than *bibliometrics* to indicate its subject.

Informetrics and quantitative science, technology and innovation (STI) studies have enforced their position as an academic discipline, so that STI indicator development is determined at least partially by an internal dynamics, although external factors play an important role as well, not in the least the business interests of large information companies. As its title indicates, the current book deals with the *application* of informetric tools. It dedicates a major part of its attention to how indicators are used in practice, and to the benefits and problems related to this use. It also discusses the relationships between the informetric domain and the research policy and management sphere, and launches proposals for new approaches in research assessment and for the development of new informetric tools.

Following Francis Narin's publication from 1976, the term *evaluative* in the book's title reflects its focus on *research assessment*. But this term refers to the *application domain* and delineates the *context* in which informetric tools are being used. It does *not* mean that informetrics is *by itself* evaluative. On the contrary, this book defends the position that informetricians should maintain *in their informetric work* a *neutral* position towards evaluative criteria or political values.

### *Target audience*

This book presents an introduction to the field of applied evaluative informetrics. It sketches the field's history, recent achievements, and its potential and limits. It also discusses the way forward both for users and for developer of informetric tools. It is written for interested scholars from all domains of science and scholarship, and especially for the following categories of readers.

- All those subjected to research assessment;
- Research students at advanced master and PhD level;
- Research managers and science policy officials;
- Research funders;
- Practitioners and students in informetrics and research assessment.

*Structure*

The book consists of six parts.

- *Part 1* presents an *introduction* to the use of informetric indicators in research assessment. It provides an historical background of the field, and presents the book's basic assumptions, main topics, structure and terminology. In addition, it includes a *synopsis*, summarizing the book's main conclusions. Readers who are interested in the main topics and conclusions of this book but who do not have the time to read it all, could focus on this part.
- *Part 2* presents an overview of the various types of *informetric indicators* for the measurement of *research performance*. It highlights the multi-dimensional nature of research performance, and presents a list of 28 often used indicators, summarizing their potential and limits. It also clarifies common misunderstandings in the interpretation of some often used statistics.
- *Part 3* discusses the *application context* of quantitative research assessment. It describes research assessment as an evaluation science, and distinguishes various assessment models. It is in this part of the book that the domain of informetrics and the policy sphere are disentangled analytically. It illustrates how external, non-informetric factors influence indicator development, and how the policy context impacts the setup of an assessment process.
- *Part 4* presents *the way forward*. It expresses the current author's views on a series of problems in the use of informetric indicators in research assessment. Next, it presents a list of new features that could be implemented in an assessment process. It highlights the potential of informetric techniques, and illustrates that *current* practices in the use of informetric indicators could be *changed*. It sketches a perspective on *altmetrics* and proposes new lines in longer term, strategic indicator research.
- *Part 5* presents five *lectures* with *historical overviews* of the field of bibliometrics and informetrics, starting from three of the field's founding fathers: Derek de Solla Price, Eugene Garfield and Francis Narin. It presents 135 slides and is based on a doctoral course presented by the author at the Sapienza University of Rome in 2015, and on lectures presented at the European Summer School of Scientometrics (ESSS) during 2010-2016, and in the CWTS Graduate Courses during 2006-2009.
- Finally, **Part 6** presents two full articles published recently by the current author in collaboration with his co-authors on hot topics of general interest in which the use of informetric indicators play a key role. These topics are: a critical comparison of five *world university rankings*; and a comparison of *usage* indicators based on the number of *full text downloads* with *citation*-based measures.

*Acknowledgements*

The author wishes to thank the following colleagues for their valuable contributions.

- Dr. Gali Halevi at The Levy Library of the Icahn School of Medicine at Mount Sinai, New York, USA for her contribution as a co-author of four articles presented in this book, on the multi-dimensional assessment of scientific research (Chapter 8); international scientific collaboration in Asia (Chapter 12); the comparison between Google Scholar and Scopus (Chapter 14); and on a comparative analysis of usage and citations (Chapter 19).
- Prof. Cinzia Daraio at the Department of Computer, Control and Management Engineering in the Sapienza University of Rome for her contribution to the text on ontology-based data base management in Section 12.3, and for her valuable comments to a draft version of Chapter 6.

- Prof. Judit Bar-Ilan at the Department of Information Science in Bar-Ilan University, Tel Aviv, Israel, for her contribution as a co-author to the paper on Google Scholar and Scopus (Chapter 14).
- The members of the Nucleo di Valutazione of the Sapienza University of Rome for stimulating discussions about the interpretation and the policy significance of world university rankings discussed in Section 10.6.

# Executive Summary

This book presents an introduction to the field of applied evaluative informetrics. Its main topic is application of informetric indicators in the assessment of research performance. It gives an overview of the field's history and recent achievements, and its potential and limits. It also discusses the way forward, proposes informetric options for future research assessment processes, and new lines for indicator development.

It is written for interested scholars from all domains of science and scholarship, especially those subjected to quantitative research assessment, research students at advanced master and PhD level, and researchers in informetrics and research assessment, and for research managers, science policy officials, research funders, and other users of informetric tools.

The use of the term informetrics reflects that the book does not only deal with bibliometric indicators based on publication and citation counts, but also with altmetrics, webometrics, and usage-based metrics derived from a variety of data sources, and does not only consider research output and impact, but also research input and process.

Research performance is conceived as a multi-dimensional concept. Key distinctions are made between publications and other forms of output, and between scientific-scholarly and societal impact. The pros and cons of 28 often used indicators are discussed.

An analytical distinction is made between *four* domains of intellectual activity in an assessment process, comprising the following activities.

- *Policy and management*: The formulation of a policy issue and assessment objectives; making *decisions* on the assessment's organizational aspects and budget. Its main outcome is a policy decision based on the outcomes from the evaluation domain.
- *Evaluation*: The specification of an evaluative framework, i.e., a set of evaluation criteria, in agreement with the policy issue and assessment objectives. The main outcome is a *judgment* on the basis of the evaluative framework and the empirical evidence collected.
- *Analytics*. Collecting, analyzing and reporting *empirical* knowledge on the subjects of assessment; the specification of an assessment *model or strategy,* and the *operationalization* of the criteria in the evaluative framework. Its main outcome is an analytical report as input for the evaluative domain.
- *Data collection*. The collection of relevant data for analytical purposes, as specified in an analytical model. Data can be either quantitative or qualitative. Its main outcome is a dataset for the calculation of all indicators specified in the analytical model.

Three *basic assumptions* of this book are the following.

- Informetric analysis is positioned in the analytics domain. A basic notion holds that from what *is* cannot be inferred what *ought to be*. Evaluation criteria and policy objectives are not informetrically demonstrable values. Of course, empirical informetric research may study quality perceptions, user satisfaction, the acceptability of policy objectives, or effects of particular policies, but they cannot provide a foundation of the validity of the quality criteria or the appropriateness of policy objectives. Informetricians should maintain in their informetric work a neutral position towards these values.
- If the tendency to replace reality with symbols and to conceive these symbols as an even a higher from of reality, are typical characteristics of *magical* thinking, jointly with the belief to be able to change reality by acting upon the symbol, one could rightly argue that the un-reflected, unconditional belief in indicators shows rather strong similarities with *magical* thinking.

- The future of research assessment lies in the intelligent *combination* of *indicators* and *peer review*. Since their emergence, and in reaction to a perceived lack of transparency in peer review processes, bibliometric indicators were used to break open peer review processes, and to stimulate peers to make the foundation and justification of their judgments more explicit. The notion of informetric indicators as a support tool in peer review processes rather than as a replacement of such processes still has a great potential.

Five strong points of the use of informetric indicators in research assessment are highlighted: it provides tools to demonstrate performance; and to shape one's communication strategies; it offers standardized approaches and independent yardsticks; it delivers comprehensive insights that reach beyond the perceptions of individual participants; and it provides tools for enlightening policy assumptions.

But severe criticisms were raised as well against these indicators. Indicators may be imperfect and *biased*; they may suggest a *façade of exactness*; most studies adopt a *limited time horizon*; indicators can be *manipulated*, and may have *constitutive effects*; measuring *societal impact* is problematic; and when they are applied, an *evaluative framework* and assessment model are often *lacking.*

The following views are expressed, partly supportive, and partly as a counter-critique towards these criticisms.

- Calculating indicators *at the level of an individual* and claiming they measure *by themselves* the individual's performance, suggests *a façade of exactness* that cannot be justified. A valid and fair assessment of individual research performance can be conducted properly only on the basis of sufficient background knowledge on the particular role they played in the research presented in their publications, and by taking into account also other types on information on their performance.
- The notion of making a *contribution to scientific-scholarly progress*, does have a basis in reality, that can best be illustrated by referring to an *historical* viewpoint. *History will show* which contributions to scholarly knowledge are valuable and sustainable. In this sense, informetric indicators do *not* measure contribution to scientific-scholarly progress, but rather indicate attention, visibility or short term impact.
- *Societal value* cannot be assessed in a politically neutral manner. The foundation of the criteria for assessing societal value is not a matter in which scientific experts have *qualitate qua* a preferred status, but should eventually take place in the policy domain. One possible option is moving away from the objective to evaluate an activity's societal *value*, towards measuring in a neutral manner researchers' *orientation* towards any articulated, lawful need in society.
- Studies on *changes in editorial and author practices* under the influence of assessment exercises are most relevant and illuminative. But the issue at stake is *not* whether scholars' practices *change* under the influence of the use of informetric indicators, but rather whether or not the application of such measures enhances *research performance.* Although this is in some cases difficult to assess without extra study, other cases clearly show traces of mere indicator manipulation with no positive effect on performance at all.
- A typical example of a constitutive effect is that research quality is more and more conceived as what citations measure. More empirical research on the size of constitutive effects is needed. If there is a genuine constitutive effect of informetric indicators in quality assessment, one should not point the critique on current assessment practices merely towards informetric indicators as such, but rather towards any claim for an absolute status of a particular way to assess research quality. Research quality is not what peers say it is, nor what citations measure.

- If the role of informetric indicators has become too dominant, it does not follow that the notion to intelligently combine peer judgments and indicators is fundamentally flawed and that indicators should be banned from the assessment arena. But it does show the combination of the two methodologies has to be organized in a more balanced manner.

- In the proper use of informetric tools an *evaluative framework and an assessment model* are indispensable. To the extent that in a practical application an evaluative framework is absent or implicit, there is *a vacuum*, that may be easily filled either with ad-hoc arguments of evaluators and policy makers, or with un-reflected assumptions underlying informetric tools. Perhaps the role of such ad hoc arguments and assumptions has nowadays become too dominant. It can be reduced only if evaluative frameworks become stronger, and more actively determine which tools are to be used, and how.

The following alternative approaches to the assessment of academic research are proposed.

- A key assumption in the assessment of academic research has been that it is not the *potential* influence or importance of research, but the *actual* influence or *impact* that is of primary *interest to policy makers* and evaluators. But an academic assessment policy is conceivable that rejects this assumption. It embodies a shift in focus from the measurement of performance itself to the assessment of *preconditions* for performance.

- Rather than using citations as indicator of research importance or quality, they could provide a tool in the assessment of *communication effectiveness*, and express the extent to which researchers bring their work to the attention of a broad, potentially interested audience. This extent can in principle be measured with informetric tools. It discourages the use of citation data as a *principal* indicator of importance.

- The *functions* of publications and other forms of scientific-scholarly output, as well as their *target audiences* should be taken into account more explicitly than they have been in the past. Scientific-scholarly journals could be systematically categorized according to their function and target audience, and separate indicators could be calculated for each category. More sophisticated indicators of internationality of communication sources can be calculated than the journal impact factor and its variants.

- One possible approach to the use of informetric indicators in research assessment is a systematic exploration of indicators as tools to set *minimum performance standards*. Using baseline indicators, researchers will most probably change their research practices as they are stimulated to meet the standards, but if the standards are appropriate and fair, this behavior will actually increase their performance and that of their institutions.

- At *the upper part* of the quality distribution, it is perhaps feasible to distinguish entities which are '*hors catégorie*', or '*at Nobel Prize level*'. Assessment processes focusing on the very top of the quality distributions could further operationalize the criteria for this qualification.

- Realistically speaking, *rankings of world universities* are here to stay. Academic institutions could, individually or collectively, seek to influence the various systems by formally sending to their creators a request to consider the implementation of a series of new features: more advanced analytical tools; more insight into how the methodological decisions influence rankings; and more information in the system about additional, relevant factors, such as teaching course language.

- In response to major criticisms towards current national research assessment exercises and performance-based funding formula, an alternative model would require less efforts, be more transparent, stimulate new research lines and reduce to some extent the Matthew Effect. The basic unit of assessment in such a model is the emerging research group rather than the individual researcher. Institutions submit emerging groups and their research programs, which are assessed in a combined peer review-based and informetric approach, applying minimum performance criteria. A funding formula is partly based on an institution's number of acknowledged emerging groups.

The practical realization of these proposals requires a large amount of informetric research and development. They constitute important elements of a wider R&D program of *applied evaluative informetrics*. The further exploration of measures of communication effectiveness, minimum performance standards, new functionalities in research information systems, and tools to facilitate alternative funding formula, should be conducted in a close collaboration between informetricians and external stakeholders, each with their own domain of expertise and responsibilities.

These activities tend to have an applied character and often a short term perspective. Strategic, longer term research projects with a great potential for research assessment are proposed as well. They put a greater emphasis on the use of techniques from *computer science* and the newly available *information and communication technologies*, and on *theoretical models* for the interpretation of indicators.

- It is proposed to develop new indicators of the *manuscript peer review process.* Although this process is considered important by publishers, editors and researchers, it still strikingly opaque. Applying classical humanistic and computational linguistic tools to peer review reports, an understanding is may be obtained for each discipline what is considered a reasonable quality threshold for publication, how it differs among journals and disciplines, and what distinguishes an acceptable paper from one that is rejected. Eventually, it could lead to better indicators of journal quality.
- To solve a series of challenges related to the management of informetric data and standardization of informetric methods and concepts, it is proposed to develop an Ontology-Based Data Management (OBDM) system for research assessment. The key idea of OBDM is to create a three-level architecture, constituted by the ontology, a conceptual, formal description of the domain of interest; the data sources; and the mapping between these two domains. Users can access the data by using the elements of the ontology. A strict separation exists between the conceptual and the logical-physical level.
- The creation is proposed of an informetric *self-assessment tool* at the level of individual authors or small research groups. A challenge is to create an online application based on key notions expressed decades ago by Eugene Garfield about author benchmarking, and by Robert K. Merton about the formation of a *reference group*. It enables authors to check the indicator data calculated about themselves, decompose the indicators' values, learn more about informetric indicators, and defend themselves against inaccurate calculation or invalid interpretation of indicators.
- As an illustration of the importance of theoretical models for the interpretation of informetric indicators, a model of a country's scientific development is presented. Categorizing national research systems in terms of *the phase* of their scientific development is a meaningful alternative to the presentation of rankings of entities based on a single indicator. In addition, it contributes to the solution of ambiguity problems in the interpretation of indicators.

It is proposed to dedicate in doctoral programs more attention to the ins and outs, potential and limits of the various assessment methodologies. Research assessment is an activity one can learn.

# Part 1. General introduction and synopsis

# 1.    Introduction

## 1.1.    Abstract

This chapter presents an introduction to the book. It starts with on overview of the value and limits the use of informetric indicators in research assessment, and presents a short history of the field. It continues with the book's main assumptions, scope and structure. Finally, it clarifies the terminology used in the book.

## 1.2.    The value and limits of informetric indicators in research assessment

What is the value of informetric indicators in the assessment of scientific-scholarly research?

- Informetric tools may help researchers and their organizations to demonstrate their performance. As an example, consider an institute developing and applying new tools in the field of ocular surgery. There is evidence that the work of the institute's staff is widely known across the globe, especially in the leading clinical centers in the field. Patients from many different countries come to the institute and undergo interventions. Specialists from all over the world seek to obtain a fellowship in the institute to learn the newest techniques. Clinical surveys confirm the success of these techniques. A way to further substantiate the contribution the institute made to research and clinical practice in its subject field is to examine traces of influence of its work in the scientific literature. Table 1.1 presents a citation analysis of the group's performance.

Table 1.1. Citation count and rank of the institute's five most frequently cited articles

| Rank | Year | Total citations | Rank in annual volume | Nr. docs in annual volume |
|------|------|-----------------|-----------------------|---------------------------|
| 1 | 1998 | 416 | 1 | 96 |
| 2 | 2006 | 312 | 1 | 267 |
| 3 | 2004 | 273 | 2 | 151 |
| 4 | 1999 | 237 | 4 | 267 |
| 5 | 1999 | 224 | 3 | 220 |

Source: Scopus. Data collected by this book's author on 28 Dec 2016. The first data row shows that the most highly cited article published from the institute in 1996 has been cited 416 times. The total number of documents published in this journal in the same year by researchers from all over the world amounts to 96. In a ranking of these 96 articles based on citation counts, the institute's paper ranks first. The table shows that all 5 articles included in the table are highly ranked in this way. A further analysis of a larger set of their highly cited work reveals that the group published both in the very specialist journals and in more general journals, and in journals with a European, British or American orientation.

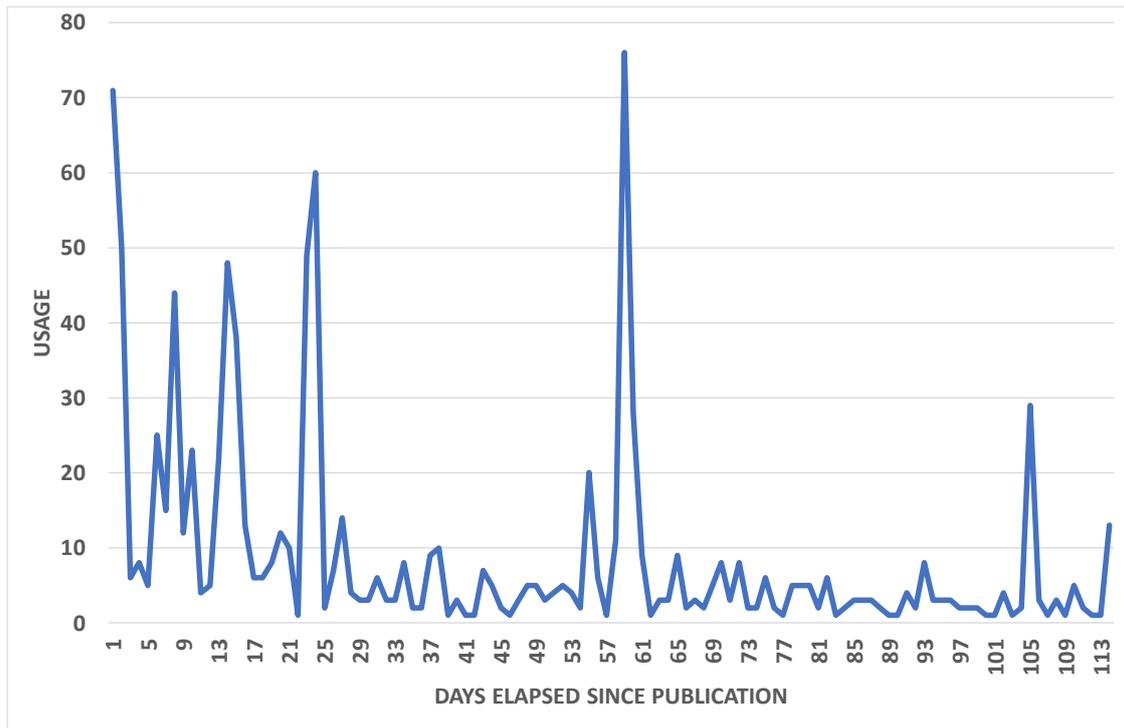

Figure 1.1. Usage counts (online views and full txt downloads) for the paper Toward new indicators of a journal's manuscript peer review process, by H.F. Moed, published in Frontiers in Research Metrics and Analytics. Usage starts soon after its online publication. The daily fluctuations during the first weeks since publication are probably due to weekend holidays. About 60 days after publication the paper was presented at the OECD Blue Sky Conference in Ghent, Belgium, 19-21 Sept. 2016. Around this date, usage has increased substantially.

- Informetric indicators can be useful tool for authors who are interested in tracking the degree of attention to their work, and in assessing the effectiveness of their communication strategies. A typical example of a useful indicator that has become available on a large scale during the past decade, is the number of online views or full text downloads of articles published in an electronic journal. Figure 1.2 presents the number of document views and downloads – denoted as usage – of one particular paper, presented at an international conference of policy makers. It shows a substantial increase in usage around the day it was presented at that conference. This case illustrates also how strongly usage counts may depend upon the behavior of the authors themselves, and therefore are, to some extent, manageable.

- The use of a well-documented and validated informetric method in an assessment process enables an evaluator to achieve a certain degree of standardization in the process, and to compare units of assessment against an independent yardstick. These characteristics are sometimes indicated with the term 'objective'. Use of such a method reduces the risk that the outcomes of an assessment are biased in favor of particular external interests. Rather than considering the collection and interpretation of indicators as an individual matter of participating evaluators, the decision to formally use in a peer review process a balanced informetric method process may contribute to an appropriate, informed outcome of the assessment.

- Informetric tools may produce insights and relationships beyond the horizon of an individual expert's knowledge. This is true not only for indicators, but especially for science maps, revealing an informetric image of innumerable relationships among objects, such as single scientific articles, authors, institutions, journals, research topics and disciplines. They may provide 'aerial views' of aggregate data reflecting the behavior of large numbers of actors, views that are complementary to the outcomes of more qualitative and disaggregated approaches.

- They can also be used as tools to critically and empirically investigate how quality perceptions are formed, and examine the validity of policy assumptions about the functioning of the scientific-scholarly system, and the effectiveness of policy measures. As argued in *Section 8.5*, informetric studies of a research system can also shed light on which type of research assessment process and which types of indicators are the most promising in a performance analysis of that system.

Informetric tools in research assessment have their *limitations* too. Severe criticism is raised against the current use of citation-based indicators and other informetric measures in research assessment. The following often debated issues are further discussed in *Chapter 9*, in which the current author gives his view and suggests possible solutions.

- *Indicators may be imperfect* or *biased.* For instance, generating attention and making a contribution to scientific-scholarly progress are not identical concepts. If one agrees that performance tends to attract attention, the degree of received attention may be interpreted as an indicator of performance. But other factors influence the level of attention as well, for instance, the way at which a piece of work is being exposed to a wide audience. The awareness that all performance indicators are 'partial' or 'imperfect'– influenced as they may be not only by the various aspects of performance but also by other factors that have little to do with performance – is as old as the use of performance indicators itself. Indicators may be *imperfect* or *biased*, but in the application of such indicators this is not seldom forgotten.

- *Most studies adopt a limited time horizon.* The time horizon in an assessment is the considered length of the time period for an objective to meet, outcome to be created or impact to be generated. Therefore, the question which *time horizon* is employed in an assessment is crucial. In most informetric assessments, the time horizon is 10 years or less, and the focus is on recent past performance, as it is believed to increase the policy relevance, and reduce data collection costs. In the calculation of altmetrics and usage counts, time windows in impact measurement can be even less than one year.

- *Indicators may be manipulated.* An often articulated critique on the use of bibliometric indicators states that if publishing articles and being cited is the norm, researchers change their behavior in order to obtain the highest possible score. This may lead to 'strategic' behavior and even to indicator manipulation. For instance, there is evidence that some journal editors seek to influence the value of journal impact factor of their journals.

- *Indicators may have constitutive effects*. Another criticism states that in the minds of those concerned the meaning of a concept that an indicator claims to measure is more and more narrowed to the definition of that indicator. Thus, scientific production is more and more equated with publishing articles, and research quality with being well cited. This phenomenon is denoted as the constitutive effect of an indicator (Dahler-Larsen, 2014).

- *Measuring societal impact is problematic.* The notion of multi-dimensional impact has created a growing interest in the societal – technological, social, economic, educational, cultural – value of research. But the time delays involved in generating impact in these extra-scientific domains may be typically 10 years or even longer. Also, societal merit cannot be measured in a politically neutral manner. What is socially valuable according to one political view, may be considered inappropriate in an alternative view.

- *An evaluative framework and assessment model are often lacking.* An evaluative framework aims to set evaluation criteria in an assessment, derived from the policy issue at stake and from the assessment objectives. Informetric indicators are often available without offering such a framework. This may make their role too dominant and give space to constitutive effects. Also, they may be applied without a well-defined assessment model, specifying how the assessment will take place, and ensuring it is not only efficient and fit-to-purpose, but also fair.

## 1.3.    A short history of bibliometrics and informetrics

### *The start*

Pioneering work has been conducted the 1960s and 1970s by Derek de Solla Price, a visionary who applied bibliometric techniques in a 'science of science', Eugene Garfield, the founder of the Science Citation Index, and Francis Narin, who introduced the term 'evaluative bibliometrics'. Chapters 13 to 17 in *Part 5* of this book give an overview of their contributions.

A theoretical basis of the use of citation-based indicators as measures of intellectual influence is provided by the notion developed in Robert K. Merton's sociological studies that references give credit where credit is due, acknowledge the community's intellectual debts to the discoverer, and can be conceived as registrations of intellectual property and peer recognition (Merton, 1957; 1996). Moed (2005a, p. 193 a.f.) gives an overview of the views of a series of authors on what cited references and citations measure. It should be emphasized that according to Merton's theory, a reference acknowledges the *source* of a knowledge claim, but not necessarily the claim's *validity*. In other words, it does not provide any theoretical justification of the claim that the more cited a piece of work is, the more valid are its results.

One of the very first studies applying citation-based indicators as measures of intellectual influence was conducted by Stephen Cole and Jonathan Cole (1967; 1971). They used these indicators as a sociological *research tool*. This type of use should be distinguished from the application of indicators in an *evaluative context* for the assessment of research performance of individuals or groups. The former type of use eventually aims at testing some kind of research *hypothesis* or revealing a structure. The latter may lead to statements on the performance of particular, designated individual scientists in the research system. As Cole (1981) puts it:

> Citations are a very good measure of the quality of scientific work for use in sociological studies of science; but because the measure is far from perfect it would be an error to reify it and use it to make individual decisions. [...] In sociological studies our goal is not to examine individuals but to examine the relationships among variables (Cole, 1989, p. 11).

The pioneering work in the USA by Narin (1976) showed that international scientific influence as measured by citations is a crucial parameter in the measurement of research performance. In Europe, Ben Martin and John Irvine further developed his approach. They proposed workable definitions of key concepts such as 'indicator', 'influence' and 'impact'.

> The citation rate is a partial indicator of the impact of a scientific publication: that is, a variable determined partly by (a) the impact on the advance of scientific knowledge, but also influenced by (b) other factors, including various social and political pressures such as the communication practices [...], the emphasis on the numbers of citations for obtaining promotion, tenure or grants, and the existing visibility of authors, their previous work, and their employing institution (Martin and Irvine, 1983, p. 70).

Anthony van Raan emphasized the *complementarity* between bibliometric indicators and peer review.

> I do not plead for a replacement of peer review by bibliometric analysis. Subjective aspects are not merely negative. In any judgment there must be room for the intuitive insights of experts. I claim, however, that for a substantial improvement of decision making an advanced bibliometric method […] has to be used in parallel with a peer–based evaluation procedure (van Raan, 2004a, p. 27).

At the end of the 1980s van Raan founded at the University of Leiden the Centre for Science and Technology Studies (CWTS), of which he was the director for almost 30 years. He established the Science and Technology Indicators Conference Series, edited the first handbook on science and technology indicators (van Raan, ed., 1987), and supervised numerous PhDs students, including the author of this book.

Tibor Braun, Wolfgang Glanzel and Andras Schubert at the Academy of Sciences in Budapest were in the early 1980s the first to systematically calculate a series of bibliometric macro-indicators derived from the Science Citation Index for all countries in the world. As from the late 1980s, many other authors started using citation and publication-based indicators in a large number of studies related to various entities: individual researchers, research groups, departments or institutes, universities, countries, journals and subject fields.

Many valuable and informative historical overviews of the field of informetrics have been published during the past decades (e.g. Mingers and Leydesdorff, 2015). A comprehensive review of citation-based indicator development is presented by Waltman (2016). Table 1.2 gives a list of the institutions of the recipients of the Derek de Solla Price Memorial Medal, periodically awarded to scientists with outstanding contributions to the fields of quantitative studies of science, by a panel comprised of the editors and members of the advisory board of the journal Scientometrics together with former Price awardees.

Table 1.2. Institutions of recipients of the Derek de Solla Price Award

| Region | Country | Institution | Year of Award |
|---|---|---|---|
| Europe | Belgium | Univ Hasselt | 2001 |
| | Belgium | Univ Antwerp | 2001 |
| | Czechoslovakia | Czechoslovak Acad Sci | 1989 |
| | Denmark | Royal Library School | 2005 |
| | France | OST/INRA | 2009 |
| | Hungary | Hungarian Acad Sciences | 1986; 1993; 1999; 2009 |
| | Netherlands | Leiden Univ | 1995; 1999 |
| | Netherlands | Univ Amsterdam | 2003 |
| | Sweden | Umea Univ | 2011 |
| | UK | City Univ London | 1989 |
| | UK | Univ. Sussex | 1997 |
| | UK | Loughborough Univ | 2015 |
| North America | USA | Inst. Scientific Inform. | 1984; 1987 |
| | USA | Univ Oregon | 1985 |
| | USA | CHI Research | 1988 |
| | USA | Columbia Univ | 1995 |
| | USA | Drexel Univ | 1997; 2005; 2007 |
| | USA | Indiana Univ | 2013 |
| Other | Israel | Bar-Ilan Univ | 2017 |
| | USSR | Moscow State Univ | 1987 |

Table 1.2 shows that the overwhelming part of the awardees were active in institutions located in European and North American countries. The medalists and their colleagues made excellent contributions to the field, and their awards are well deserved. While in the first half of the time period awardees tend to be located in the USA, in the second half they are mainly affiliated with European institutions. But Figure 1.2 reveals that institutions in other geographical regions are emerging. Both the absolute number of the percentage share of articles published in the journal Scientometrics by authors from Asia and Latin America has increased substantially over the years. In 2016, almost 40 per cent of papers was authored by an author from Asia, and about 10 per cent by a Latin American author. This outcome illustrates that an indicator based on the number of awards is to some extent conservative, and does not keep pace with the most recent developments. The bibliometric trend forecasts award winners from Asia and Latin America in the near future.

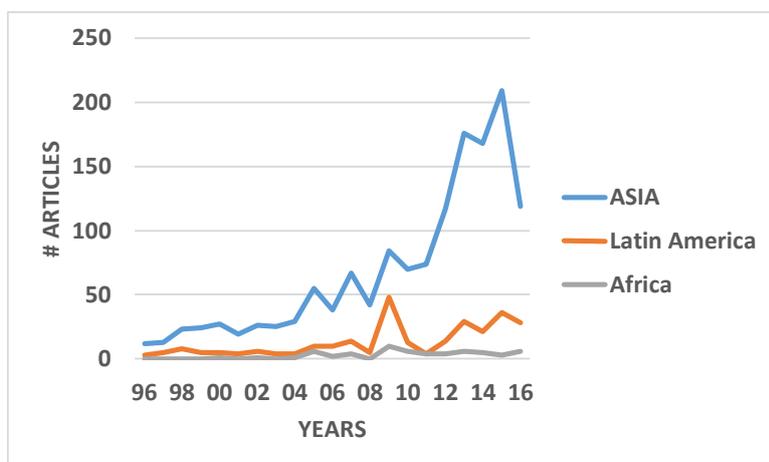

Figure 1.2. Number of articles published in the journal Scientometrics by authors from three geographical regions: Asia, Latin America and Africa. The absolute number of articles published in this journal increased over the years, from 80-100 per year during 1996-2004 to over 300 in 2016, with peaks even above 500 in 2013 and 2015. The figure shows a steady increase in the share of articles from Asia, with over 100 papers in 2016. Major publishing countries in this region are China, Taiwan, South Korea, Japan and Iran. As from 2010, there is an increase also in papers from Latin America, with 28 papers in 2016, major countries being Brazil, Mexico and Chile. The peak for this region in 2009 is probably due to the fact that in this year one of the major international conferences in the field, the Conference of the International Society for Scientometrics and Informetrics was held in Brazil. Also Africa reveals as from 2000 a publication activity in the journal. The most important country is South Africa.

***Recent developments***

In the current economical atmosphere where budgets are strained and funding is difficult to secure, ongoing, diverse and wholesome assessment is of immense importance for the progression of scientific and research programs and institutions. Research assessment is an integral part of any scientific activity. It is an ongoing process aimed at improving the quality of scientific-scholarly research. It includes evaluation of research quality and measurements of research inputs, outputs and impacts, and embraces both qualitative and quantitative methodologies, including the application of bibliometric indicators and peer review.

The manner by which research assessment is performed and executed is a key matter for a wide range of stakeholders including program directors, research administrators, policy makers and heads of scientific institutions as well as individual researchers looking for tenure, promotion or to secure funding to name a few. These stakeholders have also an increasing concern regarding the quality of research performed especially in light of competition for talent and budgets and mandated transparency and accountability demanded by overseeing bodies (Hicks, 2009).

The following trends can be identified in the *science policy domain* during the past ten years.

- *Emphasis on societal value and value for money.* In most OECD countries, there is an increasing emphasis on the effectiveness and efficiency of government-supported research. Governments need systematic evaluations for optimizing their research allocations, re-orienting their research support, rationalizing research organizations, restructuring research in particular fields, or augmenting research productivity. In view of this, they have stimulated or imposed evaluation activities of their academic institutions. Universities have become more diverse in structure and are more oriented towards economic and industrial needs.

- *Performance-based funding*. Funding of scientific research – especially in universities – tends to be based more frequently upon performance criteria, especially in countries in which research funds were in the past mainly allocated to universities by the Ministry responsible for research as a block grant, the amount of which was mainly determined by the number of enrolled students. It must be noted that in the U.S. there has never been a system of block grants for research; in this country research funding was, and is still, primarily based on peer review of the content of proposals submitted to funding organizations. Government agencies as well and funding bodies rely on evaluation scores to allocate research budgets to institutions and individuals. Such policy requires the organization of large scale research assessment exercises (OECD, 2010; Hicks, 2010) especially in terms of monetary costs, data purchasing, experts' recruitment and processing systems (Jonkers & Zacharewicz, 2016).

- *Research in a global market*. Research institutions and universities operate in a global market. International comparisons or rankings of institutions are being published on a regular basis with the aim to inform students, researchers and knowledge seeking external groups. Research managers use this information to benchmark their own institutions against their competitors (Hazelkorn, 2011). Indicators from numerous world university ranking systems such as the Shanghai, Times Higher Education, QS and Leiden Ranking, and U-Multirank play an important role.

- *Internal research assessment systems*. More and more institutions implement internal research assessment processes and build research information systems (see for instance EUROCRIS, 2013) containing a variety of relevant input and output data on the research activities within an institution, enabling managers to distribute funds based on research performance.

Due to the computerization of the research process and the digitization of scholarly communication, more and more policy-relevant data sources are becoming available. Quantitative research assessment becomes a 'big data' activity. The following specific trends can be observed.

- *Multiple, comprehensive citation indexes*. While the Science Citation Index founded by Eugene Garfield (1964) and published by the Institute for Scientific Information (currently Clarivate's Web of Science) has for many years been the only database with a comprehensive coverage of peer reviewed journals, in 2004 two new indexes entered the market, namely Elsevier's *Scopus* and *Google Scholar*.

- *Full texts in digital format*. Due to electronic publishing, more and more full texts of research publications are available in a digital format. While in the past bibliometric studies were mostly based on publication *meta-data* – including cited references –, current informetric techniques increasingly analyze *full texts*.

- *Usage data from publishers*' sites. Major publishers make their content electronically available on-line, and researchers as well as administrators are able to measure the use of their scientific output as a part of an assessment process (Luther, 2002; Bjork & Paetau, 2012).
- *Construction of large publication repositories*: Disciplinary or institutionally oriented publication repositories are being built, including meta-data and/or full text data of publications made by an international research community in a particular subject field, or by researchers active in a particular institution, respectively (Fralinger & Bull, 2013; Burns, Lana & Budd,2013).
- *Altmetric and other new data sources*: More and more researchers use social media such as Twitter, reference managers like Mendeley, and scholarly blogs, to communicate with each other and to promote their work. Traces of this use are stored in electronic files that can be analyzed with informetric techniques. Also, patent data of major patent offices are available in electronic form, and the Word Wide Web itself can be used as a data source in webometric analysis.

The above trends in the science policy domain and in the computerization of the research and communication processes generated an increasing interest in the development, availability and practical application of new indicators for research assessment.

- *Development of new indicators.* In de past decade many new indicators have been developed. They cover new dimensions of research communication and performance, and have become more sophisticated. Typical examples are altmetrics, webometrics and usage based measures, or citation-context analyses. Their development attracted more and more specialists from other disciplines, including statistical physics, molecular biology, econometrics and computer science.
- *More indicators are becoming available*. Currently, indicators such as author h-indices and total citation and publication counts are available in the three large literature databases Web of Science, Scopus and Google Scholar, or in special assessment tools such as Clarivate's Incites and Elsevier's SciVal. Many measures are produced by small, specialized firms, such as Altmetric.com or Plum Analytics, or by spin-offs such as Scimago.com and CWTS.com. Reference managers such as Mendeley and ResearchGate provide indicators as well, based on data from their own systems.
- *Desktop bibliometrics*. Calculation and interpretation of science metrics are not always made by bibliometric experts. "Desktop bibliometrics", a term coined by Katz & Hicks (1997) and referring to an evaluation practice using bibliometric data in a quick, often unreliable manner, is becoming a reality.
- *More and more critique on the use of indicators*. In a comment published in Nature, Benedictus and Miedema (2016) criticized what they denote as an "obsession with metrics" in assessment practices at Dutch academic institutions, especially academic hospitals. They argued that a pressure to publish as many papers and generate as many citations as possible created a tendency to give far too little weight to its influence of research work on patient care. Their experiences are good illustrations of the constitutive effects of indicators mentioned in the previous section.

## 1.4.    Basic assumptions

A basic notion holds that from what *is* cannot be inferred what *ought to be*. This notion has implications for the role of informetrics in the foundation of the evaluative criteria to be applied in an assessment, or of the political objectives of the assessment. Evaluation criteria and policy objectives are not informetrically demonstrable values. Of course, empirical informetric research may study quality perceptions, user satisfaction, the acceptability of policy objectives, or effects of particular policies, but they cannot provide a theoretical foundation of the validity of the quality criteria or the appropriateness of policy objectives. Informetricians should maintain in their informetric work a neutral position towards these values.

The current author conceives informetrics as a value free, empirical science. Being value free is conceived as a *methodological* requirement. This book shows how statistical definitions of indicators of research performance are based on theoretical assumptions on what constitutes performance. The informetric component and the domain of evaluative or political values in an assessment are disentangled by distinguishing between quantitative-empirical, informetric evidences on the one hand, and an evaluative framework based on normative views on what constitutes research performance and which policy objectives should be achieved, on the other. This distinction is further discussed in *Section 6.4*.

Above it was argued that informetricians should maintain *in their informetric work* a neutral position towards evaluative or political values. This statement should be further qualified in the following manner. Maintaining a neutral position towards evaluative criteria or political objectives is a methodological requirement. But informetricians are also researchers and members of the wider research community. There are *also* potential subjects of research assessment themselves. As researchers and as assessed subjects they do have views on what constitutes performance and what are appropriate and less appropriate political objectives. The methodological requirement of a value-free informetrics does not mean that they are not 'allowed' to have these views, nor that they are not allowed to communicate their views to the outside world. But they should make such views *explicit*, and make at the same time clear that these are their own views that are *not* informetrically *demonstrable*.

As regards the application of informetric tools in research assessment, a key assumption underlying this book is that the future of research assessment lies in the intelligent *combination* of *indicators* and *peer review*. From the very beginning, and in reaction to a perceived lack of transparency in peer review processes, and to the critical view of peer review as an instrument to consolidate an 'old boys' network, bibliometric indicators were used to break open peer review processes, and stimulate peers to make the foundation and justification of their judgments more explicit. The notion of informetric indicators as a support tool in peer review processes rather than as a replacement of such processes has still a great potential, and this book aims to further explore it. A necessary condition for achieving this is a thorough awareness of the potentialities and limitations of *both* methodologies.

## 1.5.    This book's main topics

This book aims to inform the ongoing debate about the validity and usefulness of informetric indicators in research assessment. It gives an overview of recent insights into their potential and limits, and does not only include *classical* bibliometric indicators based on publication and citation counts, but also newly developed measures such as *altmetrics* derived from social media appearances, and *usage* metrics based on full text downloads or online views of electronic publications.

Contrary to the situation in the 1960s, informetric tools are currently not merely used by a selected group of librarians and scientific information experts, but by many researchers in their daily practices, managers of research institutions in various types of assessment processes, politicians in the development of research funding mechanisms, the daily and weekly press in presenting the wider public detailed rankings of world universities, and information companies not only in their product portfolios but also in their marketing. The five main topics of this book are as follows.

- *An overview of new informetric tools.* A first topic is the description of a series of important *new* databases, methodologies, indicators and products introduced during the past ten years in the field of informetrics and its application in research assessment. The field has attracted interested experts from many other research disciplines, including statistical physics, social network analysis, molecular biology, and especially, computer science and artificial intelligence. And the number of new informetric tools has increased enormously. This book focuses on a series of hot topics: the use of the database Google Scholar; development phases of scientifically developing countries; new indicators derived from social media or based on full text downloads; and the publication of numerous rankings of world universities.

- *Often used informetric indicators and their pros and cons*. As outlined in the previous sections, the number of indicators available in research assessment as increased substantially over the years. A second topic is the categorization and a comprehensive overview of the most important indicators or indicator families, the aspects or dimensions they are assumed to be measuring, and their principal pros and cons. This presentation does *not* focus on the *technical* and *statistical* aspects of the indicators, but on their basic *theoretical* notions and assumptions.

- *The relationship between the informetric and the policy domain.* A third key topic in this book is a critical discussion of the current role of indicators in research assessment, and to reflect upon the relationship between the domain of informetrics and the policy domain. It is argued that, on the one hand, quantitative research assessment should not become a playing ball in the hands of the policy makers and managers using informetric tools, nor of the big and smaller information companies producing them. It is a scientific-scholarly field with its own, independent criteria of validity and good practice. On the other hand, informetric experts should not sit on the chair of a policy maker or manager, and decide –implicitly or explicitly – on normative, political issues. Instead, they should inform the policy domain, by offering instrumental support to its policies and insights and enable a reflection upon its policy objectives.

- *Options for consideration when designing an assessment process*. A fourth topic relates to the *application* of informetric indicators. The book addresses a series of basic issues and proposes a series of *options* that could be considered when designing and implementing a research assessment process. The list of options aims to illustrate the potential of informetric techniques. It aims to create openness and space for further reflection by illustrating that *current* practices in the use of informetric indicators could be *changed*. In this way it shows what *could* be done, *not* what *should* be done. As argued above, the latter issue is to be solved in the definition of an evaluative framework, integrating policy needs and informetric evidences with a view of what is valuable and must be achieved. Informetricians, including the author of this book, should maintain a neutral position towards the normative aspects of such a framework.

- *Future research and indicator development*. A last topic relates to the development of new indicators for research assessment. The book proposes new lines of research that can facilitate the development of new indicators. Two main components are distinguished. The first is an emphasis on the need to develop theoretical models as guides for interpretation and adequate use of indicators. During the past decades, *data* collection and handling have been a key priority. But nowadays it becomes clear more data does not always mean a better understanding. Such models can in principle come from all interested disciplines. A second component is an emphasis on the use of methods from computer science, and of the potential of the new information and communication technologies. Perhaps many indicators that are currently still being applied, as well as the way in which they are used, are determined by techniques that were developed some 50 years ago. New technical and analytical approaches can bring more progress than they have achieved thus far.

## 1.6.    Structure of book

The book consists of six parts.

- **Part 1** presents an introduction to the use of informetric indicators in research assessment. It provides an historical background to the topic, and presents the book's basic assumptions, main topics, structure and terminology. In addition, *Chapter 2* presents a synopsis, summarizing the book's main conclusions of each of its 25 chapters in about 1,000 words.

- **Part 2** presents an overview of the use of informetric *indicators for the measurement of research performance*. While *Chapter 3* focuses on the multi-dimensional nature of research performance, *Chapter 4* presents a list of 28 informetric indicators (or indicator families) that are often used as measures of research performance, and summarizes their pros and cons. *Chapter 5* discusses two common misunderstandings, namely that journal impact factors are good predictors of the citation rate of individual articles, and that in large data sets errors or biases are always canceled out. In addition, it dedicates attention to the interpretation of particular often used statistics, namely (linear) correlation coefficients between two variables.

- **Part 3** discusses the *application context* of quantitative research assessment. *Chapter 6* describes research assessment as an evaluation science. It is in this chapter that the domain of informetrics and the policy sphere are disentangled analytically. The distinctions function as a framework for an outline presented in *Chapters 7 and 8* of how external, non-informetric factors influence indicator development. While *Chapter 7* discusses the influence of *non- or extra-informetric factors*, relating to evaluative assumptions, the wider social context, and business interests, *Chapter 8* focuses on the *policy context*. It introduces the notion of the multi-dimensional research assessment matrix and that of meta-analyses generating background insights for policy makes and evaluators responsible for an assessment.

- **Part 4** presents *the way forward*. It starts in *Chapter 9* with a discussion of a series of major problems in the use of informetric indicators in research assessment: the assessment of individual scholars; the use of a limited time horizon; the assessment of societal impact; the effects of the use of indicators upon authors and editors, and their constitutive effects; and the need for an evaluative framework and an assessment model. The author of this book expresses his views on these problems, and on how they could be dealt with. This chapter forms an introduction to Chapters 10, 11 and 12 discussing the way forward.

- The way forward in *quantitative research assessment* is the subject of *Chapter 10*. It presents a list of new features that could be implemented in an assessment process. They highlight the potential of informetric techniques, and illustrate that *current* practices in the use of informetric indicators could be *changed*. *Chapter 11* sketches a perspective on *altmetrics*, also termed as 'alternative' metrics, but many propositions and suggestions relate to the use in research assessment of any type of informetric indicator. This chapter, as well as *Chapter 12* propose new lines in indicator research that put a greater emphasis on *theoretical models*, and on the use of techniques from *computer science* and the newly available *information and communication technologies*. They advocate the development of models of scientific development, and new indicators of the manuscript peer review process, ontology-based data management systems, and informetric author self-assessment tools.

- **Part 5** presents five *lectures* with *historical overviews* of the field of bibliometrics and informetrics, starting from three of the field's founding fathers: Derek de Solla Price, Eugene Garfield and Francis Narin. It is based on a doctoral course presented by the author at the Sapienza University of Rome in 2015, and on lectures presented at the European Summerschool of Scientometrics (ESSS) during 2010-2016, and in the CWTS Graduate Courses during 2006-2009. Main topics addressed are: citation networks; science mapping; informetric databases; the science-technology interface; journal metrics; and research performance indicators.

- Finally, **Part 6** presents two full articles published recently by the author of this book on hot topics of general interest in which the use of informetric indicators play a key role. These topics are: A critical comparison of five *world university ranking* systems; and how *usage indicators* based on the number of full text downloads or online reads of research articles compare with citation-based measures. These articles provide background information on the chapters presented in *Part 4.*

## 1.7.    A note on terminology

This book uses the term 'informetrics' as a generic term indicating the study of quantitative aspects of information. It comprises all studies denoted as 'bibliometric ', including the classical publication- and citation-based studies, but it is broader, as it does not merely analyze books and other media of written communication, but also altmetric and usage data, webometric, economic and research input data, and survey data on scholarly reputation.

The term *assessment* is used as an *overarching* concept, denoting the total of activities in assessment or evaluation processes, or the act of evaluating or assessing *in general*. This book uses the term evaluation exclusively in relation to the setting of criteria for, and the formation of judgments on, the 'worth' of a subject. This use is further explained in Chapter 6 below. It can be said that an assessment process often contains an evaluative component, but there may be assessment without such a component. This book avoids the use of the term 'evaluation' as a noun, but uses its adjective form in combination with nouns such as 'framework' or 'criteria'.

This book focuses on *academic* research, primarily intended to increase scholarly knowledge, but often motivated by and funded for specific technological objectives such as the development of new technologies such as medical breakthroughs. It comprises both 'curiosity-driven' or 'pure' as well as 'strategic' or 'application oriented' research. The latter type of research may be fundamental in nature, but is undertaken in a quest for a particular application, even though its precise details are not yet known. Indicators (also denoted synonymously as metrics throughout this book) are conceived as instruments used to measure the various components of research activity.

# 2.    Synopsis

## 2.1.    Abstract

This chapter provides summaries of the main topics and conclusions of each chapter.

## 2.2.    Part 1. Introduction

### *Chapter 1*

Chapter 1 highlights five strong points of the use of informetric indicators in research assessment. It helps to demonstrate one's performance; it gives information for shaping one's communication strategies; it offers standardized approaches and independent yardsticks; it delivers comprehensive insights; and provides tools for enlightening policy assumptions.

Five main points of critique are: Indicators may be biased and not measure what they are supposed to measure; most studies adopt a limited time horizon; indicators can be manipulated, and may have constitutive effects; measuring societal impact is problematic; and when they are used, an evaluative framework and assessment model are often lacking.

The chapter describes a series of trends during the past decade in the domain of science policy: an increasing emphasis on societal value and value for money, performance-based funding and on globalization of academic research, and a growing need for internal research assessment and research information systems.

Due to the computerization of the research process and the digitization of scholarly communication, research assessment is more and more becoming a 'big data' activity, involving multiple comprehensive citation indexes, electronic full text databases, large publications repositories, usage data from publishers' sites, and altmetric, webometric and other new data sources.

The above trends created an increasing interest in the development, availability and application of new indicators for research assessment. Many new indicators were developed, and more and more mearues have become available on a large scale. Desktop bibliometrics is becoming a common assessment practice.

Two principal assumptions of the author are as follows.

- From what is cannot be inferred what ought to be. Evaluation criteria and policy objectives are not informetrically demonstrable values. Informetric research may study such values empirically, but cannot provide a theoretical foundation of the validity of the quality criteria or the appropriateness of policy objectives. Informetricians should in their informetric work maintain a neutral position towards these values.
- The future of research assessment lies in the intelligent combination of indicators and peer review. From the very beginning, bibliometric indicators stimulated peers to make the foundation and of their judgments more explicit. The notion of informetric indicators as a support tool in peer review processes rather than as a replacement of such processes still has a great potential, and this book aims to further explore it. A necessary condition for achieving this is a thorough awareness of the potentialities and limitations of both methodologies.

The main subjects of the book are:

- An overview of important new databases, methodologies, indicators and products introduced during the past 10 years in the field of informetrics and its application in research assessment.
- A comprehensive overview of the most important indicators or indicator families, the aspects or dimensions they are assumed to measure, and their potential and limits.
- A clarification of the relationship between the informetric, the evaluative and the policy domain.
- Possible new features that could be implemented in a research assessment process.
- New lines of research that are expected to lead to the development of new, useful indicators.

This book uses the term '*informetrics*' as a generic term indicating the study of quantitative aspects of information. It does not only analyze bibliometric data based on publication and citation counts, but also altmetric and usage data, webometric data, economic data, research input data, and survey data on scholarly reputation.

The term '*assessment'* is used as an *overarching* concept, denoting the total of activities in assessment or evaluation processes, or the act of evaluating or assessing *in general*. This book uses the term evaluation exclusively in relation to the setting of criteria for, and the formation of judgments on, the 'worth' of a subject.

This book focuses on *academic* research, primarily intended to increase scholarly knowledge, but often motivated by and funded for specific technological objectives such as the development of new technologies such as medical breakthroughs. It comprises both 'curiosity-driven' or 'pure' as well as 'strategic' or 'application oriented' research.

## 2.3.    Part 2. Informetric indicators of research performance

### Chapter 3

The multi-dimensional nature of research performance is highlighted. Four main components of research activity are distinguished: input, process, output and impact. *Input* includes funding, manpower and research infrastructure. Indicators of research infrastructure are not primarily performance indicators but relate rather to a *precondition* for performance. *Process* indicators focus on research collaboration and efficiency.

*Output* can be publication based, such as a journal article or monograph, or be delivered in a non-publication format, such as research dataset, and be directed to the scientific-scholarly community or to society and the wider public. A key distinction is made between scientific-scholarly and societal *impact*. Societal impact embraces a wide spectrum of aspects outside the domain of science and scholarship itself, including technological, social, economic, educational and cultural aspects.

Research performance is reflected in *all four* components. For instance, research funding, especially competitive funding, can be assumed to depend on the past performance and the reputation of the grant applicant, and therefore relates both to input and to impact. Efficiency is both a process and an impact indicator, as it aims to measure output or impact per unit of input. A table is presented that lists the 28 important indicators or indicator families and their pros and cons.

### Chapter 4

**Chapter 4** presents the main characteristics of the most important indicator families.

- *Publication-based indicators.* In academic institutions, publications constitute in all scientific-scholarly subject fields an important format of academic output. But in the private sector, and also in academic departments with a strong applied orientation, primarily aiming to produce new products or processes, publishing in the public, peer reviewed literature often does not have the highest priority; in this case, other output forms must be considered as well. Publication counts may be used to define minimum performance standards.

- *Citation-based indicators.* Citation analysis offers a certain degree of standardization, and compares units of assessment against an independent yardstick, which makes an evaluator more independent from the views of the subjects of the analysis and of the commissioning entity. Citations can be interpreted as proxies of more direct measurements of intellectual influence, but they are by no means indicators of the validity of a knowledge claim. Citation impact and quality do not coincide; and, as all indicators, may be affected by disturbing factors and suffer from serious biases.

- *Journal metrics.* The quality or impact of the journals in which a unit under assessment has published is a performance aspect in its own right. But the relationship between a journal's impact factor and the rigorousness of its manuscript peer review process is unclear. Journal metrics cannot be used as a surrogate of actual citation impact; they are no good predictors of the citation rate of individual papers. Moreover, their values can to some extent be manipulated, and may be affected by editorial policies.

- *Patent-based indicators.* Analyses of inventors of the findings described in patents reveal the extent to which scientists with academic positions contributed to technological developments. Patent-to-patent citations may reveal a patent's technological value, and patent citations to scientific papers a technology's science base. But the propensity to apply for patents differs across countries because of legislation or culture, and also across subject fields. In addition, patents are a very poor indicator of the commercialization of research results.

- *Usage-based indicators.* Data on downloads of an electronic publication in html or full text format enable researchers to assess the effectiveness of their communication strategies, and may reveal attention of scholarly audiences from other research domains or of non-scholarly audiences. Downloaded articles may be selected according to their face value rather than their value perceived after reflection. Also, there is an incomplete data availability across providers, and counts can be manipulated. It is difficult to ascertain whether downloaded publications were actually read or used.

- *Altmetrics* relate to different types of data sources with different functions. Mentions in social media may reveal impact upon non-scholarly audiences, and provide tools to link scientific expertise to societal needs, but cannot be used to measure scientific-scholarly impact. Their numbers can be manipulated, and interdependence of the various social media may boost figures. Readership counts in scholarly reference managers are potentially faster predictors of emerging scholarly trends than citations are, but results depend upon readers' cognitive and professional background.

- *Webometrics.* Indicators of Web presence and impact are extracted from a huge universe of documents available on the Web. They do not merely relate to institutions' research mission, but also to their teaching and social service activity. But there is no systematic information on the universe of Web sources covered and their quality; and an institution's Web presence depends upon its internal policies towards the use of the Web, and upon the propensity of its staff to communicate via the Web.

- *Economic indicators.* Indicators of economic value and efficiency are relevant measures in research assessment. But not all contributions to economic development can be easily measured. The relationship between input and output is not necessarily linear, and may involve a time delay. Accurate, standardized input data is often unavailable; and comparisons across countries are difficult to make, due to differences in the classification of economic data. Indicators of funding from industry are useful measures of economic value; but funding levels differ across disciplines, and data may be difficult to collect.

- *Reputation and esteem based measures*. Receiving a prestigious prize or award is a clear manifestation of esteem. But absolute numbers tend to be low; the evaluation processes on which the nominations are based are not always fully transparent; and there are no agreed equivalences that apply internationally and facilitate comparison across disciplines. Reputation can be measured in surveys using validated methods from social sciences. But response rates are often very low; mentions may be based on 'hear-say' rather than on founded judgement and may refer to performance made in a distant past.

- *Measures of scientific collaboration, migration and cross-disciplinarity*. Data on co-authorship and on how authors migrate over time from one institutional affiliations to another, provide useful indicators of intra-institutional, national, international scientific collaboration and migration. But instances that have not resulted in publications remain bibliometrically invisible. Indicators of cross-disciplinary measure the relevance of a piece of research for surrounding disciplines, or the cognitive breadth of research impact. Their calculation presupposes a valid, operational classification of research into disciplines.

- *Indicators of research infrastructure* are not primarily performance indicators, but focus on preconditions for performance. They measure basic facilities that support research, the scale of the research activities, and their sustainability. But research practices differ across disciplines; large research teams or laboratories are mostly found in the natural and life sciences. It is difficult to obtain reliable, comparable institutional data, as there is no agreement on the basis of a full cost calculation of research investment. There is no clear, generally accepted definition of being research active across universities, countries and disciplines.

The calculation of informetric indicators of research performance more and more becomes a *'big data'* activity. Not only the increasing volume of informetric datasets is of interest, but especially their *combination* creates a large number of new possibilities. For instance, the combination on an article-by-article basis of citation indexes and usage log files of full text publication archives, enables one to investigate the relationships between downloads and citations, and develop ways to generate a more comprehensive, multi-dimensional view of the impact of publications than each of the sources can achieve individually.

There is an increasing interest in mapping techniques, and science mapping is to be qualified as one of the most important domains of informetrics as a big data science. It can be defined as the development and application of computational techniques for the visualization, analysis, and modeling of a broad range of scientific and technological activities as a whole.

### *Chapter 5*

Although this book does not present details on the technical and statistical aspects of most informetric indicators, **Chapter 5** does discuss three common misunderstandings as regards interpretation of particular often used statistical measures or techniques, related to journal impact factors as means of skewed distributions, errors in large datasets, and the interpretation of linear correlation coefficients. The conclusions are as follows.

- Journal impact factors are *no* good predictors of the citation rate of individual articles in a journal.
- Only random errors tend to cancel out in large datasets; systematic biases may remain.
- When interpreting a correlation coefficient, never rely merely on its numerical value. Consider always a scatter plot of the underlying data.

## 2.4.  Part 3. The application context

### *Chapter 6*

In ***Chapter 6*** an analytical distinction is made between *four* domains of intellectual activity in an assessment process, including the following activities.

- *Policy or management*: The formulation of a policy issue and assessment objectives; making *decisions* on the assessment's organizational aspects and budget. Its main outcome is a policy decision based on the outcomes from the evaluation domain.
- *Evaluation*: A specification of the evaluative framework, i.e., a set of evaluation criteria in agreement with the constituent policy issue and assessment objectives. The main outcome is a *judgment* on the basis of the evaluative framework and the empirical evidence collected.
- *Analytics*. Collecting, analyzing, and reporting *empirical* knowledge on the subjects of assessment; The specification of an assessment *model or strategy,* and the *operationalization* of the criteria in the evaluative framework. Its main outcome is an analytical report as input for the evaluative domain.
- *Data collection*. Collection of relevant data for analytical purposes, as specified in the analytical model. Data can be either quantitative or qualitative. Its main outcome is a dataset for the calculation of all indicators specified in the analytical model.

A main objective of this analytical categorization is to distinguish between scientific-methodological principles and considerations on the one hand, and policy-related, political or managerial considerations on the other. Focusing on the role of informetricians, the chapter argues as follows.

- What is of worth, good, desirable, or important in relation to the functioning of a subject under assessment, *cannot* be established in informetric, or, more general, in quantitative-empirical research. The principal reason is that one cannot infer what *ought to be* from what actually *is*.
- What informetric investigators *can* do is empirically examine value *perceptions* of researchers, the conditions under which they were formed and the functions they fulfil, but the foundation of the validity of a value is *not* a task of quantitative-empirical, informetric research. The same is true for the formation of *evaluative judgements*. The latter two activities belong to the domain of evaluation.
- Informetricians should maintain a *methodologically* neutral position towards the constituent policy issue, the criteria specified in the evaluative framework, and the goals and objectives of the assessed subject. As professional experts, their competence lies *primarily* in the development and application of analytical models *given* the established evaluative framework.
- Obviously, informetricians are allowed to form and express 'normative' views while assessing a unit's worth, but when doing so they should make these explicit and give them methodologically speaking a hypothetical status.

Several types of assessment models are distinguished: peer review based versus indicator based assessments, and self-assessment versus external assessment. It distinguishes four classes of *evaluation strategies*:

- Scientific-experimental focusing on impartiality and objectivity;
- Management-oriented systems based on systems theory;
- Qualitative anthropological approaches emphasizing the importance of observation and space for subjective judgement;
- Participant oriented strategies, giving a central role to 'consumers'.

A genuine challenge is to combine the various models and create hybrid assessment models.

### Chapter 7

**Chapter 7** illuminates the influence of non-informetric or extra-informetric factors on the development of indicators, and in this way aims to *disentangle* informetric arguments and evaluative principles, one of the key objectives of this book. Typical examples are given as regards the following issues: size dependent versus size independent indicators; focus on the top or the bottom of a performance distribution; indicator normalizations; definition of benchmark sets; and the application of a short term or a long term perspective.

For instance, a series of citation impact indicators seek an 'optimal' combination of publication and citation counts, and address the issue whether this optimum should be size-normalized or not. Under the surface of this seemingly technical debate, a confrontation takes place of distinct views of what constitutes genuine research performance or 'quality'.

- According to one view, a citation-per-paper ratio is the best indicator, because it helps to detect 'saturation', which occurs if a research group increases its annual number of published papers while the citation impact per paper declines.
- A second view holds that such a ratio penalizes groups with a large publication output, while a large publication output should be rewarded rather than penalized.
- A third view aims to reduce the role of absolute publication numbers by proposing an indicator counting only the 'best' papers in terms of citation impact.
- A fourth view claims that the only good performance indicator is an efficiency indicator dividing output or impact by 'input' measures.

It illustrates how in seemingly technical discussions on the construction and statistical properties of science indicators, *'evaluative', theoretical assumptions on what constitutes research performance* play an important, though often implicit, role. Such values are denoted as extra-informetric, as their validity cannot be grounded in informetric research.

Chapter 7 also presents a brief history of some of the main lines in bibliometric indicator development from the early 1960s up to date. It focuses on the *wider socio-political context* in which indicators were developed. It describes the context of their launch, not so much in terms of the *intentions* of the developers, but, at a higher analytical level, in terms of how they fit into – or are the expression of – a more general tendency in the policy, political or cultural environment in which they were developed. A base assumption is that knowledge of the wider context in which specific indicators were developed contributes to a better understanding of how and under which conditions they can be properly applied.

It is argued that in the early decades newly proposed indicators primarily aimed to solve specific policy problems and fitted in with specific national or institutional policy contexts, but during the past 10-15 years the following two tendencies emerged: on the one hand, previously developed indicators were used in more and more policy contexts, including contexts in which they are only partially or hardly fit-for-purpose; on the other hand, indicator development was more and more driven by an internal dynamics powered by mathematical-statistical considerations.

Finally, Chapter 7 discusses the influence of *business interests* of the information industry upon the development of indicators. It concludes that since Eugene Garfield introduced the Journal Impact Factor as an 'objective' tool to expand the journal coverage of his citation index independently of journal publishers, the landscape of scientific information providers and users has changed significantly. While, on the one hand, politicians and research managers at various institutional levels need valid and reliable fit-for-purpose metrics in the assessment of publicly funded research, there is, on the other hand, a tendency that indicators increasingly become a tool in the business strategy of companies with product portfolios that may include underlying databases, social networking sites, or even indicator products. This may be true both for 'classical' bibliometric indicators and for alternative metrics.

### Chapter 8

*Chapter 8* argues that in the design of a research assessment process, one has to decide which methodology should be used, which indicators to calculate, and which data to collect. To make proper decisions about these matters, one should address the following key questions, each of which relates to a particular aspect of the research assessment process.

- What is the unit of the assessment? A country, an institution, a research group, an individual, or a research field or an international network? In which discipline(s) is it active?
- Which dimension of the research process must be assessed? Scientific-scholarly impact? Social benefit? Multi-disciplinary? Participation in international networks?
- What are the purpose and the objectives of the assessment? Allocate funding? Improve performance? Increase regional engagement?
- What are relevant, general or 'systemic' characteristics of the units of assessment? For instance, to which extent are they oriented towards the international research front?

The answers to these question determine which indicators are the most appropriate in a particular assessment process. Indicators that are useful in one context, may be less so in another. A warning is issued against a practice in which particular indicators are used in a given context simply because there are available, and because they have been successfully applied in *other* contexts.

Knowledge on general characteristics of the system of units of assessment plays an important role in the formulation of the objectives of an assessment. Such assumptions do not focus on *individual* units, but on more general or systemic characteristics of these units *as a group*. Therefore, they can be denoted as 'meta' assumptions, and illuminate the assessment's *policy context.*

For instance, if an analysis of the state of a country's science system provides evidence that researchers tend to publish mainly in national journals without a serious manuscript peer review process, it is from an informetric viewpoint *defensible* to use the number of publications in the top quartile of journals in terms of citation impact as an indicator of research performance, not so much as an *evaluation tool*, but rather as an instrument to *change* certain communication *practices* among researchers.

However, if in internationally oriented, leading universities one has to assess candidates submitting their job application, it is questionable whether it makes sense comparing applicants according to the average citation impact of the journals in which they published their papers. Due to *self-selection*, the applicants will probably publish at least a large part of the papers in good, international journals, and in this group journal impact factors hardly discriminate between high and lower performance.

## 2.5.    Part 4. The way forward

### Chapter 9

This chapter discusses a series of problems in the use of informetric indicators for evaluative purposes. Its main conclusions are as follows. Their implications for the application of indicators and for future indicator development are further discussed in Chapters 10-12.

- *The problem of assessing individual scholars.* Calculating indicators at the level of an individual and claiming they measure *by themselves* the individual's performance suggests a façade of exactness that cannot be justified. A valid and fair assessment of individual research performance can be conducted properly only on the basis of sufficient background knowledge on the particular role they played in the research presented in their publications, and by taking into account also other types on information on their performance.
- *The effect of a limited time horizon*. The notion of making a contribution to scientific-scholarly progress, does have a basis in reality, that can best be illustrated by referring to an *historical* viewpoint. *History will show* which contributions to scholarly knowledge are valuable and sustainable. In this sense, informetric indicators do *not* measure contribution to scientific-scholarly progress, but rather indicate attention, visibility or short term impact.
- *The problem of assessing societal impact*. Societal value cannot be assessed in a politically neutral manner. The foundation of the criteria for assessing societal value is not a matter in which scientific experts have *qualitate qua* a preferred status, but should eventually take place in the policy domain. One possible option is moving away from the objective to evaluate an activity's societal *value*, towards measuring in a neutral manner researchers' *orientation* towards *any* articulated, lawful need in society, as reflected for instance in *professional contacts.* Due to time delays, emphasis on societal impact on the one hand, and assessment focus on *recent* past performance on the other, are at least partially conflicting policy incentives.
- *The effects of the use of indicators upon authors and editors.* Studies on changes in editorial and author practices under the influence of assessment exercises are most relevant and illuminative. The issue at stake is *not* whether scholars' practices *change* under the influence of the use of informetric indicators, but rather whether or not the application of such measures enhances their *research performance.* Although this is in some cases difficult to assess without extra study, other cases clearly show traces of mere indicator manipulation with no positive effect on performance at all.
- *How to deal with constitutive effects of indicators.* A typical example of a constitute effect is when research quality is more and more perceived as what citation measure. If the tendency to replace reality with symbols and to conceive these symbols as an even a higher from of reality, are typical characteristics of *magical* thinking, jointly with the belief to be able to change reality by acting upon the symbol, one could rightly argue that the un-reflected, unconditional belief in indicators shows rather strong similarities with *magical* thinking.

- More empirical research on the size of constitutive effects is urgently needed. If there is a genuine constitutive effect of informetric indicators in quality assessment at all, one should not point the critique on current assessment practices merely towards informetric indicators as such, but rather towards any claim for an *absolute status* of a particular *way* to assess quality. If the role of informetric indicators has become too dominant, it does *not* follow that the idea to intelligently combine peer judgements and indicators is fundamentally flawed and that indicators should be banned from the assessment arena. But it does show the combination of the two methodologies has to be organized in a more sophisticated and balanced manner.

- *The need for an evaluative framework and an assessment model*. Chapter 6 underlines the need to define an evaluative framework and an assessment model. To the extent that in a practical application an evaluative framework is absent or implicit, there is a vacuum, that may be easily filled either with ad-hoc arguments of evaluators and policy makers, or with un-reflected assumptions underlying informetric tools. Perhaps the role of such ad hoc arguments and assumptions has nowadays become too dominant. It can be reduced only if evaluative frameworks become stronger, and more actively determine which tools are to be used, and how. To facilitate this, informetricians should make the assumptions and hidden normative a-priories of their tools explicit, and inform policy makers and evaluators about the potential and the limits of these tools.

**Chapter 10**

Chapter 10 critically reflects on the assumptions underlying current practices in the use of informetric indicators in research assessment, and proposes a series of alternative approaches, indicating their pros and cons.

*Communication effectiveness as a precondition for performance.*

In academic institutions, especially in research universities, it is considered appropriate to stimulate that academic researchers make a solid contribution to scientific-scholarly progress. But is it defensible to require that they generate impact? What should be of primary interest to academic policy makers: importance (potential influence) or impact (actual influence)? An academic assessment policy is conceivable that rejects the claim that impact rather than importance is the key aspect to be assessed, and discourages the use of citation data as a *principal* indicator of importance.

Such an assessment process would *not* aim at measuring importance or *contribution to scientific-scholarly progress* as such, but rather *communication effectiveness*, a concept that relates to a *precondition for* performance rather than to performance itself. It expresses the extent to which researchers bring their work to the attention of a broad, potentially interested audience, and can in principle be measured with informetric tools.

*Some new indicators of multi-dimensional research output*

The *functions* of publications and other forms of scientific-scholarly output, as well as their *target audiences* should be taken into account more explicitly than they have been in the past. Journals with an educative or enlightening function are important in scientific scholarly communication, and tend to have a substantial societal value. Since their visibility at the international research front as reflected in citations and journal impact factors may be low, in a standard bibliometric analysis based on publication and citation counts this value may not be visible.

Scientific-scholarly journals could be systematically categorized according to their function and target audience, and separate indicators could be calculated for each category. In an analysis of research output in journals directed towards *national* audiences, citation-based indicators are *less relevant*. At the same time, in citation analyses based on the large international citation indexes focusing on the international research front, it would be appropriate to disregard such journals. It is proposed to develop indicators of *journal internationality* based on the geographical distribution of publishing, citing or cited authors. A case study shows that journal impact and internationality are by no means identical concepts. Whether or not a journal is indexed in one or more of the large citation indexes is not in all cases a good indicator of its international orientation.

*Definition of minimum performance standards*

One possible approach to the use of informetric indicators in research assessment is a systematic exploration of indicators as tools to set minimum performance standards and define in this way a performance baseline. Important considerations in favor of this approach are as follows.

- There is evidence that citation rates are a good predictor of how peers discriminate between a 'valuable' and a 'less valuable' past performance, but that they do not properly predict within the class of 'valuable' performances, peers' perception of 'genuine excellence' This outcome underlines the potential of informetric indicators in the assessment of the *lower part* of the quality distribution.
- Using indicators to define a baseline, researchers will most probably change their research practices as they are stimulated to meet the standards, but if the standards are appropriate and fair, this behavior will actually increase their performance and that of their institutions.
- Focusing on minimum criteria involves a shift in perspective from measuring performance *as such* towards assessing the *preconditions* for performance. Expert opinion and background knowledge will play a crucial role, not only in the definition of the standards themselves, but also in the assessment processes in which these are applied.
- The definition of minimum standards could also be applied to journal impact measures. Rather than focusing on the most highly cited journals and rewarding publications in this top set, it would be possible to discourage publication in the bottom set of journals (e.g., the bottom quartile) with the lowest citation impact.

At the *upper part* of the quality distribution, it is perhaps feasible to distinguish entities which are '*hors categorie*', or '*at Nobel Prize level*'. Assessment processes focusing on the *top* of the quality distributions could further operationalize the criteria for this qualification.

*Policy towards World University Rankings*

*Chapter 18* in *Part 6* of this book presents a comparative analysis of 5 World University Ranking Systems. Realistically speaking, rankings of world universities are here to stay. When university managers use their institution's position in world university rankings primarily for marketing purposes, they should not disregard the negative effects such use may have upon researchers' practices within their institution. They should also critically address the validity of the methodological claims made by producers of these ranking systems.

The following strategy towards these ranking systems is proposed. Academic institutions could, individually or collectively, seek to influence the various systems by formally sending them a request to consider the implementation of the following new features.

- Offer more advanced analytical tools, enabling a user to analyze the data in a more sophisticated manner than ranking systems currently offer.

- Provide more insight into how the methodological decisions of the producers influence the ranking positions of given universities.
- Enhance the information in the system about additional factors, such as teaching course language.

In addition, academic institutions could proceed as follows.

- Create a special university webpage providing information on a university's internal assessment and funding policies, and on its various types of performance, and giving comments on the methodologies and outcomes of ranking systems.
- Request ranking producers to make these pages directly accessible via their systems.

*An alternative approach to performance based funding*

Major criticisms towards national research assessment exercises underline their bureaucratic burden, costs, lack of transparency, and their Matthew effect. Adopting an informetric viewpoint, an alternative assessment model is described that would require less efforts, be more transparent, stimulate new research lines, and reduce to some extent the Matthew Effect. The main features are as follows.

- The base unit of assessment is an *emerging group*, a small research group with a great scientific potential. Acknowledged as a 'rising star', the director has developed a promising research program, and has already been able to establish a small research unit.
- The profile of an emerging group should be further operationalized into a set of minimum quantitative criteria, taking into account the communication practices and funding opportunities in the group's subject field.
- In the assessment procedure, institutions submit groups rather than individual staff. Submissions provide information on the group's past performance and a future research programme, that are evaluated in a peer review process, informed by appropriate informetric indicators.
- The primary aim of the peer review is to define the minimum standards in operational terms and assess whether the submitted groups comply with these standards. These standards constitute a precondition for future to the group's future performance.
- There would be no need to rank groups, assign ratings, discriminate between 'top' and 'almost top' groups, or make funding decisions. Funding decisions take place within their institution.
- To stimulate the implementation of quality control processes within an institution, the availability of a certain amount of funding from internal, performance-based allocation processes could be posed as a necessary condition.
- A part of public funding (block grant) could be allocated to institutions as a lump sum on the basis of the number of acknowledged emerging groups.

The practical realization of these proposals requires a large amount of informetric research and development. They constitute important elements of a wider R&D program in *applied evaluative informetrics*. These activities tend to have an applied character and often a short term perspective, and focus on the development side of R&D. They should be conducted in a close collaboration between informetricians and external stakeholders. Chapters 11 and 12 propose strategic, longer term research projects with a great potential for research assessment.

### Chapter 11

Chapter 11 discusses the potential of altmetrics. A multi-dimensional conception of altmetrics is proposed, namely as traces of the computerization of the research process, and conceived as a tool for the practical realization of the ethos of science and scholarship in a computerized or digital age. Three drivers of development of the field of altmetrics are distinguished.

- In the policy domain: An increasing awareness of the multi-dimensionality of research performance, and an emphasis on societal merit.
- In the domain of technology: The development of information and communication technologies (ICTs), especially social media.
- From the scientific-scholarly community itself: The Open Science movement.

Four aspects of the computerization of the research process are highlighted: the computerization of research data collection and analysis; the digitization of scientific information; the use of computerized communication technologies; and informetrization of research assessment.

Michael Nielsen's set of creative ideas constitute a framework in which altmetrics can be positioned. He argued that "to amplify cognitive intelligence, we should scale up collaborations, increasing cognitive diversity and the range of available expertise as much as possible". The role of altmetrics and other informetric indicators would not merely be, passively, to provide descriptors, but also actively, or proactively, to establish and optimize, Nielsen's "architecture of attention", a configuration that combines the efforts of researchers and technicians on the one hand, and the wider public and the policy domain on the other, and that "directs each participant's attention where it is best suited—i.e., where they have maximal competitive advantage".

It should not be overlooked that a series of distinctions made in 'classical' research assessment are most relevant in connection with altmetrics as well: scientific-scholarly versus societal impact; attention versus influence; opinion versus scientific fact; peer-reviewed versus non-peer reviewed; intended or unintended versus constitutive effects of indicators.

### Chapter 12

Chapter 12 proposes a series of alternative approaches to the development of new informetric indicators for research assessment that put a greater emphasis on *theoretical models* for the interpretation of informetric indicators, and on the use of techniques from *computer science* and the newly available *information and communication technologies*.

*Towards new indicators of the manuscript peer review process*

A proposal to develop new indicators of the manuscript peer review process is based on the following considerations.

- Manuscript peer review is considered important by journal publishers, editors and researchers. But the process itself is still strikingly opaque.
- Reviewers tend to receive little training in what is one of the key academic activities, and there is little evidence of any standardization in how review reports are composed.
- There is little systematic, objective information on the quality of the process across journals and subjects, and on its effect upon the quality of submitted papers.
- With peer review largely a black box, proxies for its quality have grown up, most notably the journal impact factor (JIF) based on citation counts.

- But the digitization of scientific information offers great potential for the development of tools to allow peer review to be analyzed directly.

The objectives and set-up of the project are as follows.

- The aim of the project is to build up for each discipline an understanding of what is considered a reasonable quality threshold for publication, how it differs among journals and disciplines, and what distinguishes an acceptable paper from one that is rejected.
- The analysis consists of two phases, an explorative phase in which a classical-humanities approach is dominant, aimed to develop a conceptual model; and a data mining phase, applying techniques from digital humanities.
- Taking into account a journal's scope and instructions to reviewers, the various elements of a review report should be analyzed. Statements are categorized in terms of aspect and modality. Standards and a-priories applied by a reviewer are identified.
- Relevant concepts and their indicators are developed, including the formative content of a review report, and the distance a reviewer maintains towards his own methodological and theoretical views.

The project could have the following outcomes

- It provides insight into the effects and 'added value' of peer review upon manuscript quality
- It defines a set of minimum quality standards per journal and per discipline. This information enhances the transparency of the review process, and is useful both for editors, reviewers and authors.
- It proposes indicators characterizing the various aspects of the process, for instance, its formative effect, and other tools to monitor the process.
- The results may be used to improve the quality of the peer review process.
- Ultimately, perhaps journal-level metrics can be validated that can supersede proxies such as journal impact factors.

*Towards an ontology-based informetric data management system*

During the past decades, development and application of informetric indicators in research assessment have pose a series of challenges to the management – collection, handling, integration, analysis and maintenance – of informetric data, and in the design of S&T indicators. There are data-related, concept-related and maintenance-relate issues.

To solve these issues, it is proposed to develop an Ontology-Based Data Management (OBDM) system for research assessment, along the lines set out in Daraio et al (2016). The key idea of OBDM is to create a three-level architecture, constituted by a) the ontology; b) the data sources; and c) the mapping between the two.

An *ontology* can be defined as a conceptual, formal description of the domain of interest to a given entity (e.g., organization or community of users), expressed in terms of relevant concepts, *attributes* of concepts, *relationships* between concepts, and *logical* assertions characterizing the domain knowledge. The *sources* are the data repositories accessible by the organization in which data concerning the domain are stored. The mapping is a precise specification of the correspondence between the data contained in the *data sources* on the one hand, and the elements of the *ontology* on the other.

The main advantages of an OBDM approach are as follows

- Users can access the data by using the elements of the ontology. A strict separation exists between the conceptual and the logical-physical level.
- By making the representation of the domain explicit, the acquired knowledge can be easily re-used.
- The mapping layer explicitly specifies the relationships between the domain concepts in the ontology and the data sources. It is useful also for documentation and standardization purposes.
- The system is more flexible. It is for instance not necessary to merge and integrate all the data sources at once, which could be extremely time consuming and costly.
- The system can be more easily extended. New elements in the ontology or data sources can be added incrementally when they become available. In this sense, the system is dynamical and develops over time.

*Towards informetric self-assessment tools*

The creation is proposed of an informetric self-assessment tool at the level of individual authors or small research groups. Such an application would be highly useful, but is currently unavailable. A challenge is to make an optimal use of the potentialities of the current information and communication technologies and create an online application based on key notions expressed decades ago by Eugene Garfield about author benchmarking, and by Robert K. Merton about the formation of a *reference group*.

The general concept is as follows.

- In a first step, the application enables an author to define a set of his publications he/she wishes to take into account in the assessment. It is important that there is a proper data verification tool at hand.
- In a next step, a benchmark set is created of authors with whom the assessed author can best be compared, along the lines adopted by Eugene Garfield in his proposal for an algorithm to create for a given author under assessment a set of 'candidate' benchmark authors who have bibliometric characteristics that are similar to those of the given author.
- Garfield's idea could be further developed by creating a flexible benchmarking feature as the practical realization of Robert K. Merton's notion of a reference group, i.e., the group with which individuals compare themselves, but to which they do not necessarily belong but aspire to.
- The calculated indicators should be the result of simple statistical operations on absolute numbers. Not only the outcome, but also the underlying numbers themselves should be visible.
- In addition, researchers must have the opportunity to decompose and reconstruct indicators. It should also be possible to insert particular data manually.

An adequate assessment of individual research performance can take place only by taking into account multiple sources of information about their performance. This does not mean that bibliometric measures in the assessment of individuals are irrelevant, especially when used as self-assessment tools. The proposed tool would be useful for the following reasons.

- In their self-assessments, researchers may wish to calculate specific fit-for-purpose indicators that are not 'standard' and therefore unavailable at the websites of indicator producers. The proposed self-assessment tool could be a genuine alternative to using journal impact factors or h-indices.
- Even if one is against the use of informetric indicators in individual assessments, one cannot ignore their availability to a wider public. Therefore, it would be useful if researchers had an online application to check the indicator data calculated about them, and to decompose the indicators' values.

- In this way they would learn more about the ins and outs of evaluative informetrics, and, for instance, become aware of how the outcomes of an assessment depend upon the way benchmark sets are being defined. This would enable researchers to defend themselves against inaccurate calculation or invalid interpretation of indicators.

*Towards informetric models of scientific development*

Scientifically developing countries need tools to monitor the effectiveness of their research policies in a framework that categorizes national research systems in terms of *the phase* of their scientific development. Leaving out the dynamical aspects of a system gives an incomplete picture. The current section presents a model of a country's scientific development using bibliometric indicators based on publications in international, peer-reviewed journals.

The model aims to provide a framework in which the use of informetric indicators of developing countries makes sense, as an alternative to common bibliometric rankings based on publication and citation counts from which the only signal is that such countries tend to feature in the bottom.

A simplified and experimental bibliometric model for different phases of development of a national research system distinguishes four phases:

- *Pre-development phase*. Low research activity without clear policy of structural funding of research
- *Building up*. Collaborations with developed countries are established; national researchers enter international scientific networks
- *Consolidation and expansion*. The country develops its own infrastructure; the amount of funds available for research increases
- *Internationalization.* Research institutions in the country start as fully-fledged partners, increasingly take the lead in international collaboration

The distinction into phases is purely *analytical* and does not imply a c*hronological order*. The model assumes that during the various phases of a country's scientific development, the number of published articles in peer-reviewed journals shows a more or less continuous increase, although the rate of increase may vary substantially over the years and between countries.

It is the share of a country's internationally co-authored articles that discriminates between the various phases in the development. The model also illustrates the *ambiguity* of this indicator, as a high percentage of internationally co-authored papers at a certain point in time may indicate that a country is either in the building up or the internationalization phase.

The model is applied to empirical data on South-East Asian countries, and on Arab Gulf states and neighboring countries in the Middle East, and provided evidence, for instance, that while Saudi Arabia is in the building-up phase, Iran has already entered the internationalization phase.

## 2.6. Part 5. Lectures

### Chapter 13

**Chapter 13** presents two visionary papers published by Derek de Solla Price, the founding father of the science of science. It presents his view on the scientific literature as a network of scientific papers, and introduces important informetric concepts, including research front and immediacy effect.

Next, the chapter shows how his pioneering work on modelling the relational structure of subject space evolved into the creation of a series of currently available, advanced science mapping tools.

### Chapter 14

A comparative analysis of three big, multi-disciplinary citation indexes is presented in **Chapter 14**. It starts with a presentation of the basic principles of the Science Citation Index (SCI, later Thomson Reuters' Web of Science, currently Clarivate Analytics), a multi-disciplinary citation index created by Eugene Garfield in the early 1960s. Next, it presents a study conducted in 2009 comparing the Web of Science with Scopus, a comprehensive citation index launched by Elsevier in 2004, and a recent study comparing Scopus with an even more comprehensive citation index, Google Scholar, also launched in 2004.

### Chapter 15

**Chapter 15** discusses studies on the relationship between science and technology. It starts with presenting the pioneering work by Francis Narin and co-workings on the citation analysis of the linkage between science and technology. Next, it discusses several theoretical models of the relationship between science and technology. As an illustration of an analysis of the development of a technological field, it presents key findings from a study on industrial robots. The chapter ends with illustrating the limitations of citation analysis of the scientific literature for the measurement of technological performance.

### Chapter 16

**Chapter 16** deals with journal metrics. It starts with a discussion of the journal impact factor, probably the most well-known bibliometric measure. It shows some of its technical limitations and dedicates in an analysis of editorial self-citations special attention to its sensitivity to manipulation. Next, a series of alternative journal citation measures is presented, SJR, Eigenfactor, SNIP, CiteScore, and indicators based on usage.

### Chapter 17

Definitions and properties of a series of informetric indicators discussed in earlier chapters of this book, are presented in **Chapter 17**: relative citation rates, h-index, Integrated Impact Indicator, usage-based indicators, social media mentions, and research efficiency or productivity measures. It highlights their potential and limits, and gives typical examples of their application in research assessment.

## 2.7.    Part 6. Papers

### Chapter 18

To provide users insight into the value and limits of world university rankings, **Chapter 18** presents a comparative analysis of 5 World University Ranking Systems: ARWU, the Academic Ranking of World Universities, also indicated as 'Shanghai Ranking'; The Leiden Ranking created by the Centre for Science and Technology Studies (CWTS); The Times Higher Education (THE) World University Ranking; QS World University Rankings; and an information system denoted as U-Multirank created by a consortium of European research groups.

As all ranking systems claim to measure essentially academic excellence, one would expect to find a substantial degree of consistency among them. The overarching issue addressed in this chapter is the assessment of this consistency-between-systems. To the extent that a lack of consistency is found, what are the main causes of the observed discrepancies? A series of analyses is presented, from which the following conclusions were drawn.

- Each ranking system has its proper orientation or 'profile'; there is no 'perfect' system. There is only a limited overlap between the top 100 segments of the 5 rankings.
- What appears in the top of a ranking depends to a large extent upon a system's geographical coverage, rating methodologies applied, indicators selected and indicator normalizations carried out.
- Current ranking systems are still one-dimensional in the sense that they provide finalized, seemingly unrelated indicator values rather than offer a dataset and tools to observe patterns in multi-faceted data.
- To enhance the level of understanding and adequacy of interpretation of a system's outcomes, more insight is to be provided to users into the methodological differences between the various systems, especially on how their institutional coverage, rating methods, the selection of indicators and their normalizations influence the ranking positions of given institutions.

***Chapter 19***

A statistical analysis of full text downloads of articles in Elsevier's ScienceDirect covering all scholarly disciplines reveals large differences between disciplines, journals, and document types as regards their download frequencies, their skewness, and their correlation with Scopus-based citation counts. Download counts tend to be two orders of magnitude higher and less skewedly distributed than citations. Differences between journals are discipline-specific.

Despite the fact that in all study journals download and citation counts per article positively correlate, the following factors differentiate between downloads and citations.

- *Usage leak*. Not all full text downloads of a publisher archive's documents may be recorded in the archive's log files.
- *Citation leak*. Not all relevant sources of citations may be covered by the database in which citations are counted.
- *Downloading* the full text of a document does not necessarily mean that it is fully *read.*
- *Reading and citing populations may be different*. For instance, industrial researchers may read scientific papers but not cite them as they do not publish papers themselves.
- *Number of downloads depends upon type of document*. For instance, editorials and news items may be heavily downloaded but poorly cited compared to full length articles.
- *Downloads and citations show different obsolescence functions*. Download and citation counts both vary over time, but in a different manner, showing different maturing and decline rates.
- *Downloads and citations measure different aspects*. Short term downloads tend to measure readers' awareness or attention whereas citations result from authors' reflection upon relevance.
- *Downloads and citations may influence one another in multiple ways*. More downloads may lead to more citations. But the reverse may be true as well. Articles may gain attention and be downloaded because they are cited.
- *Download counts are more sensitive to manipulation*. While citations tend to be regulated by the peer review process, download counts are more sensitive to manipulation.

- *Citations are public, usage is private*. While citations in research articles in the open, peer reviewed literature are public acts, downloading documents from publication archives is essentially a private act.